\documentclass[runningheads]{llncs}
\usepackage{graphicx}
\usepackage{color}
\newcommand{\todo}[1]{}
\renewcommand{\todo}[1]{{\color{red} ({#1})}}
\newcommand{\pat}[1]{}
\renewcommand{\pat}[1]{{\color{blue} (Pat:\ {#1})}}

\usepackage{cite}
\usepackage{listings}
\usepackage{booktabs}
\usepackage[caption=false]{subfig}
\usepackage{makecell}

\usepackage[firstpage]{draftwatermark}
\SetWatermarkText{\hspace*{6in}\raisebox{7.15in}{\includegraphics[scale=0.1]{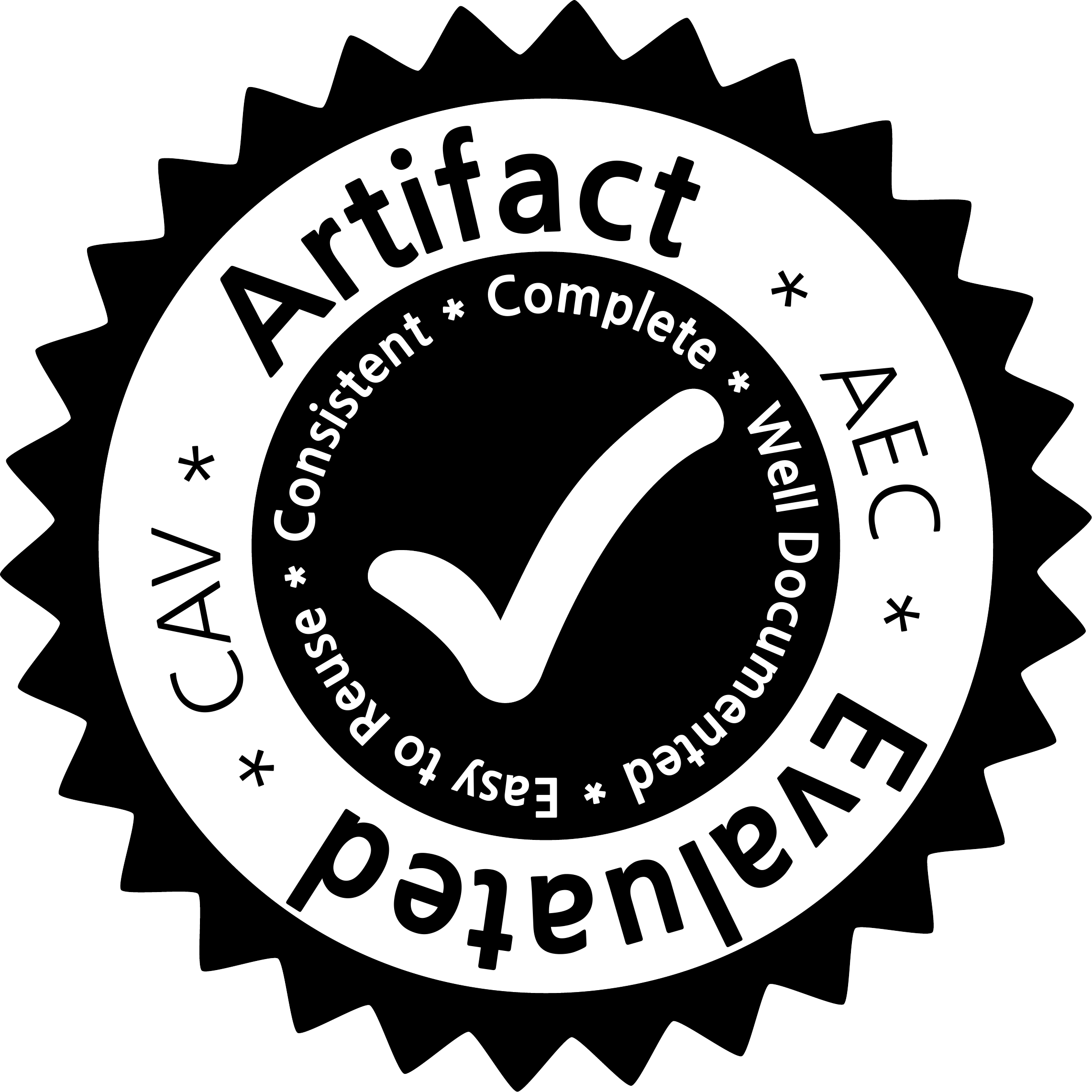}}}
\SetWatermarkAngle{0}

\def\CC{{C\nolinebreak[4]\hspace{-.05em}\raisebox{.4ex}{\tiny\bf ++}}}

\definecolor{dkgreen}{rgb}{0,0.6,0}
\definecolor{gray}{rgb}{0.5,0.5,0.5}
\definecolor{mauve}{rgb}{0.58,0,0.82}
\definecolor{myblue}{rgb}{0.36,.34,.42}
\begin{document}
%
\title{fault: A Python Embedded Domain-Specific Language For Metaprogramming Portable Hardware Verification Components}
\titlerunning{fault: Python Hardware Verification DSL}
%
\author{Lenny Truong \and
Steven Herbst \and
Rajsekhar Setaluri \and
Makai Mann \and
Ross Daly \and
Keyi Zhang \and
Caleb Donovick \and
Daniel Stanley \and
Mark Horowitz \and
Clark Barrett \and
Pat Hanrahan}
\authorrunning{L. Truong et al.}
%
\institute{Stanford University, Stanford CA 94305, USA}
\maketitle              
\begin{abstract}

While hardware generators have drastically improved design productivity, they have introduced new challenges for the task of verification.  To effectively cover the functionality of a sophisticated generator, verification engineers require tools that provide the flexibility of metaprogramming.  However, flexibility alone is not enough; components must also be portable in order to encourage the proliferation of verification libraries as well as enable new methodologies. This paper introduces \textbf{fault}, a Python embedded hardware verification language that aims to empower design teams to realize the full potential of generators.

\end{abstract}

\section{Introduction}
The new golden age of computer architecture relies on advances in the design and implementation of computer-aided design (CAD) tools that enhance productivity~\cite{10.1145/3282307, DBLP:conf/snapl/TruongH19}.  While hardware generators have become much more powerful in recent years, the capabilities of verification tools have not improved at the same pace~\cite{lockhart2018tpu}.  This paper introduces \textbf{fault},\footnote{\url{https://github.com/leonardt/fault}} a domain-specific language (DSL) that aims to enable the construction of flexible and portable verification components, thus helping to realize the full potential of hardware generators.  


Using flexible hardware generators~\cite{10.1109/MM.2010.81,bachrach:2012:chisel} drastically improves the productivity of the hardware design process, but simultaneously increases verification cost. A \emph{generator} is a program that consumes a set of parameters and produces a hardware module.
The scope of the verification task grows with the capabilities of the generator, since more sophisticated generators can produce hardware with varying interfaces and behavior.  To reduce the cost of attaining functional coverage of a generator, verification components must be as flexible as their design counterparts. To achieve flexibility, hardware verification languages must provide the metaprogramming facilities found in hardware construction languages~\cite{bachrach:2012:chisel}.

However, flexibility alone is not enough to match the power of generators; verification tools must also enable the construction of portable components.  Generators facilitate the development of hardware libraries and promote the integration of components from external sources.   Underlying the utility of these libraries is the ability for components to be reused in a diverse set of environments.  The dominance of commercial hardware verification tools with strict licensing requirements presents a challenge in the development of portable verification components.  To encourage the proliferation of verification libraries, hardware verification languages must design for portability across verification tools.  Design for portability will also promote innovation in tools by simplifying the adoption of new technologies, as well as enable new verification methodologies based on unified interfaces to multiple technologies.

This paper presents \textbf{fault}, a domain-specific language (DSL) embedded in Python designed to enable the flexible construction of portable verification components. As an embedded DSL, \textbf{fault} users can employ all of Python's rich metaprogramming capabilities in the description of verification components.  
Integration with \textbf{magma}~\cite{magma}, a hardware construction language embedded in Python, is an essential feature of \textbf{fault} that enables full introspection of the hardware circuit under test.
By using a staged metaprogramming architecture, \textbf{fault} verification components are portable across a wide variety of open-source and commercial verification tools.  A key benefit of this architecture is the ability to provide a unified interface to constrained random and formal verification, enabling engineers to reuse the same component in simulation and model checking environments.  \textbf{fault} is actively used by academic and industrial teams to verify digital, mixed-signal, and analog designs for use in research and production chips.  This paper demonstrates \textbf{fault}'s capabilities by evaluating the runtime performance of different tools on a variety of applications ranging in complexity from unit tests of a single module to integration tests of a complex design.  These experiments leverage \textbf{fault}'s portability by reusing the same source input across separate trials for each target tool.

\section{Design}
\begin{figure}
    \centering
    \includegraphics[width=\textwidth]{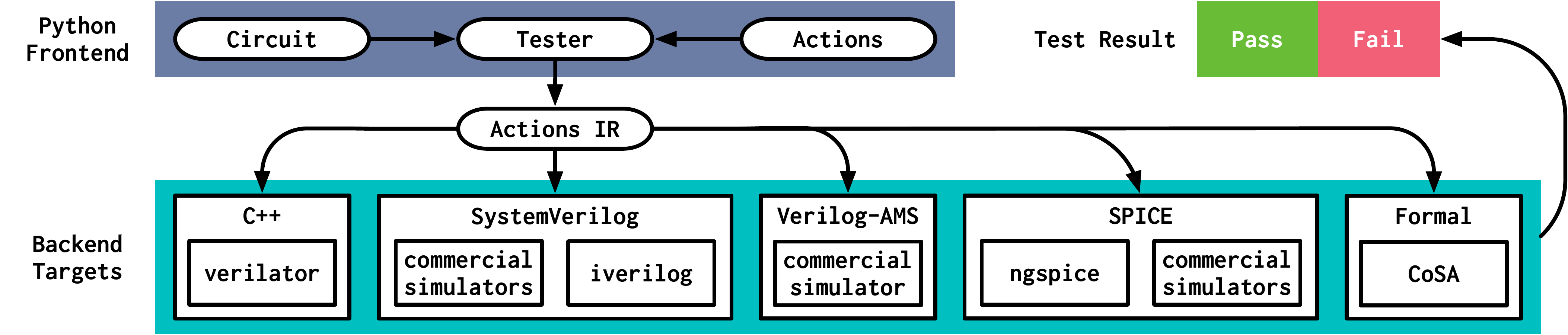}
    \caption{Architectural overview of the \textbf{fault} testing system.  In a Python program, the user constructs a \texttt{Tester} object with a \textbf{magma} \texttt{Circuit} and records a sequence of test \texttt{Actions}.  The compiler uses the action sequence as an intermediate representation (IR).  Backend targets lower the actions IR into a format compatible with the corresponding tool and provide an API to run the test and report the results.}
    \label{fig:system-diagram}
\end{figure}

We had three goals in designing \textbf{fault}: enable the construction of flexible test components through metaprogramming, provide portable abstractions that allow test component reuse across multiple target environments, and support direct integration with standard programming language features.  The ability to metaprogram test components is a vital requirement for scaling verification efforts to cover the space of functionality utilized by hardware generators.  
Portability widens the target audience of a reusable component and enhances a design team's productivity by enabling simple migration to different technologies.
Integration with a programming language enables design teams to leverage standard software patterns for reuse as well as feature-rich test automation frameworks.

Figure~\ref{fig:system-diagram} provides an overview of the system architecture.  \textbf{fault} is a DSL embedded in Python, a prolific dynamic language with rich support for metaprogramming and a large ecosystem of libraries.  
\textbf{fault} is designed to work with \textbf{magma}~\cite{magma}, a Python embedded hardware construction language which represents circuits as introspectable Python objects containing ports, connections, and instances of other circuits.  
While \textbf{fault} and \textbf{magma} separate the concerns of design and verification into separate DSLs, they are embedded in the same host language for simple interoperability.  This multi-language design avoids the complexity of specifying and implementing a single general purpose language without sacrificing the benefits of tightly integrating design and verification code.

To construct \textbf{fault} test components, the user first instantiates a \texttt{Tester} object with a \textbf{magma} circuit as an argument.  The user then records a sequence of test actions using an API provided by the \texttt{Tester} class.  Here is an example of constructing a test for a 16-bit \texttt{Add} circuit:
\begin{center}
\begin{tabular}{c}
\begin{lstlisting}[basicstyle=\ttfamily\scriptsize]
tester = Tester(Add16)
tester.poke(Add16.in0, 3)
tester.poke(Add16.in1, 2)
tester.eval()
tester.expect(Add16.out, 5)
\end{lstlisting}
\end{tabular}
\end{center}
The \texttt{poke} action (method) sets an input value, the \texttt{eval} action triggers evaluation of the circuit (the effects of \texttt{poke} actions are not propagated until an \texttt{eval} action occurs), and the \texttt{expect} action asserts the value of an output.  Attributes of the \texttt{Add16} object refer to circuit ports by name.

\textbf{fault}'s design is based on the concept of staged metaprogramming~\cite{taha2000metaml}; the user writes a program that constructs another program to be executed in a subsequent stage.  In \textbf{fault}, the first stage executes Python code to construct a test specification; the second stage invokes a target runtime that executes this specification.  To run the test for the 16-bit \texttt{Add}, the user simply calls a method and provides the desired target:
\begin{center}
\begin{tabular}{c}
\begin{lstlisting}[basicstyle=\ttfamily\scriptsize]
tester.compile_and_run("verilator")
tester.compile_and_run("system-verilog", simulator="iverilog")
\end{lstlisting}
\end{tabular}
\end{center}

By applying staged metaprogramming, \textbf{fault} allows the user to leverage the full capabilities of the Python host language in the programmatic construction of test components.  For example, a test can use a native \texttt{for} loop to construct a sequence of actions using the built-in random number library and integer type:
\begin{center}
\begin{tabular}{c}
\begin{lstlisting}[basicstyle=\ttfamily\scriptsize]
for _ in range(32):
    N = (1 << 16) - 1
    in0, in1 = random.randint(0, N), random.randint(0, N)
    tester.poke(Add16.in0, in0)
    tester.poke(Add16.in1, in1)
    tester.eval()
    tester.expect(Add16.out, (in0 + in1) & N)
\end{lstlisting}
\end{tabular}
\end{center}
Python \texttt{for} loops are executed during the first stage of computation and are effectively ``unrolled'' into a flat sequence of actions.  Other control structures such as \texttt{while} loops, \texttt{if} statements, and function calls are handled similarly.

Python's object introspection capabilities greatly enhance the flexibility of \textbf{fault} tests.  For example, the core logic of the above test can be generalized to support an arbitrary width \texttt{Add} circuit by inspecting the interface:
\begin{center}
\begin{tabular}{c}
\begin{lstlisting}[basicstyle=\ttfamily\scriptsize]
# compute max value based on port width (length)
N = (1 << len(Add.in0)) - 1
in0, in1 = random.randint(0, N), random.randint(0, N)
tester.poke(Add.in0, in0)
tester.poke(Add.in1, in1)
tester.eval()
tester.expect(Add.out, (in0 + in1) & N)
\end{lstlisting}
\end{tabular}
\end{center}
This ability to metaprogram components as a function of the design under test is an essential aspect of \textbf{fault}'s design.  It allows the construction of generic components that can be reused across designs with varying interfaces and behavior.

\textbf{fault}'s embedding in Python's class system provides an opportunity for reuse through inheritance.  For example, a design team could subclass the generic  \texttt{Tester} class and add a new method to perform an asynchronous reset sequence:
\begin{center}
\begin{tabular}{c}
\begin{lstlisting}[basicstyle=\ttfamily\scriptsize]
class ResetTester(Tester):
    def __init__(self, circuit, clock, reset_port):
        super().__init__(self, circuit, clock)
        self.reset_port = reset_port
        
    def reset(self):
        # asynchronous reset, negative edge
        self.poke(self.reset_port, 1)
        self.eval()
        self.poke(self.reset_port, 0)
        self.eval()
        self.poke(self.reset_port, 1)
        self.eval()
\end{lstlisting}
\end{tabular}
\end{center}

Combining inheritance with introspection, we can augment the the \texttt{ResetTester} to automatically discover the reset port by inspecting port types:
\begin{center}
\begin{tabular}{c}
\begin{lstlisting}[basicstyle=\ttfamily\scriptsize]
class AutoResetTester(ResetTester):
    def __init__(self, circuit, clock):
        # iterate over interface to find reset (assumes exactly one)
        for port in circuit.interface.ports.values():
            if isinstance(port, AsyncResetN):
                reset_port = port
        super().__init__(self, circuit, clock, reset_port)
\end{lstlisting}
\end{tabular}
\end{center}

\subsection{Frontend: Tester API} \label{sec:design-frontend} 

\textbf{fault}'s Python embedding is implemented by the \texttt{Tester} class which provides various interfaces for recording test actions as well as methods for compiling and running tests using a specific target.  By using Python's class system to perform a shallow embedding~\cite{boulton1992experience}, \textbf{fault} avoids the complexity of processing abstract syntax trees and simply uses Python's standard execution to construct test components.  As a result, programming in \textbf{fault} is much like programming with a standard Python library.  This design choice reduces the overhead of learning the DSL and simplifies aspects of implementation such as error messages, but comes at the cost of limited capabilities for describing control flow.  The \textbf{fault} frontend described in this paper focuses on implementation simplicity, but the system is designed to be easily extended with new frontends using alternative embeddings.


\subsubsection{Action Methods}
The \texttt{Tester} class provides a low-level interface for recording actions using methods.  
The basic action methods are \texttt{poke} (set a port to a value), \texttt{expect} (assert a port equals a value), \texttt{step} (invert the value of the clock), \texttt{peek} (read the value of a port), and \texttt{eval} (evaluate the circuit).  The \texttt{peek} method returns an object containing a reference to the value of a circuit port in the current simulation state.  Using logical and arithmetic operators, the user can construct expressions with this object and pass the result to other actions. For example, to expect that the value of the port \texttt{O0} is equal to the inverse of the value of port \texttt{O1}, the user would write \texttt{tester.expect(circuit.O0, $\sim$tester.peek(circuit.O1))}.
The \texttt{Tester} provides a \texttt{print} action to display simulation runtime information included the peeked values.  



\subsubsection{Metaprogramming Control Flow} Notably absent from the basic method interface described above are control flow abstractions.  As noted before, standard Python control structures such as loops and \texttt{if} statements are executed in the first stage of computation as part of the metaprogram.  However, there are cases where the user intends to preserve the control structure in the generated code, such as long-running loops that should not be unrolled at compile time or loops that are conditioned on dynamic values from the circuit state.  For example, consider a \texttt{while} loop that executes until it receives a ready signal:
\begin{center}
\begin{tabular}{c}
\begin{lstlisting}
     # Construct while loop conditioned on circuit.ready.
     loop = tester._while(tester.peek(circuit.ready))
     loop.expect(circuit.ready, 0)  # executes inside loop
     loop.step(2)                            # executes inside loop
     # Check final state after loop has exited
     tester.expect(circuit.count, expected_cycle_count)
\end{lstlisting}
\end{tabular}
\end{center}

This logic could not be encoded in the metaprogram, because the metaprogram is evaluated before the test is run, and thus does not know anything about the runtime state of the circuit.  To capture this dynamic control flow, the \texttt{Tester} provides methods for inserting \texttt{if-else} statements, \texttt{for} loops, and \texttt{while} loops.  Each of these methods returns a new instance of the current \texttt{Tester} object which provides the same API, allowing the user to record actions corresponding to the body of the control construct.  The \texttt{Tester} class provides convenience functions for using these control structures to generate common patterns, such as \texttt{wait\_on}, \texttt{wait\_until\_low}, and \texttt{wait\_until\_posedge}.

\subsubsection{Attribute Interface}  While the low-level method interface is useful for writing complex metaprograms, simple components are rather verbose to construct.  To simplify the handling of basic actions like \texttt{poke} and \texttt{peek}, the \texttt{Tester} object exposes an interface for referring to circuit ports and internal signals using Python's object attribute syntax.  For example, to poke the input port \texttt{I} of a circuit with value \texttt{1}, one would write \texttt{tester.circuit.I = 1}.  This interface supports referring to internal signals using a hierarchical syntax.  For example, referring to port \texttt{Q} of an instance \texttt{ff} can be done with \texttt{tester.circuit.ff.Q}.


\subsubsection{Assume/Guarantee}\label{sec:assume-guarantee}  The \texttt{Tester} object provides methods for specifying assumptions and guarantees that are abstracted over constrained random and formal model checking runtime environments.  An \emph{assumption} is a constraint on input values, and a \emph{guarantee} is an assertion on output values.  Assumptions and guarantees are specified using Python \texttt{lambda} functions that return symbolic expressions referring to the input and output and ports of a circuit.  For example, the guarantee  \texttt{lambda a, b, c: (c >= a) and (c >= b)} states that the output \texttt{c} is always greater than or equal to the inputs \texttt{a} and \texttt{b}. Here is an example of verifying a simple ALU using the assume/guarantee interface:
\begin{center}
    \begin{tabular}{c}
\begin{lstlisting}
# Configuration sequence for opcode register
tester.circuit.opcode_en = 1
tester.circuit.opcode = 0  # opcode for add (+)
tester.step(2)
tester.circuit.opcode_en = 0
tester.step(2)
# Verify add does not overflow
tester.circuit.a.assume(lambda a: a < BitVector[16](32768))
tester.circuit.b.assume(lambda b: b < BitVector[16](32768))
tester.circuit.c.guarantee(
    lambda a, b, c: (c >= a) and (c >= b)
)
\end{lstlisting}
\end{tabular}
\end{center}
Note that this example demonstrates the use of \texttt{poke} and \texttt{step} to initialize circuits not only for constrained random testing, but also for formal verification.





\subsection{Actions IR} \label{sec:design-ir}
In using the \texttt{Tester} API, users construct a sequence of \texttt{Action} objects that are used as an intermediate representation (IR) for the compiler.  Basic port action objects, such as \texttt{Poke} and \texttt{Expect}, simply store references to ports and values.  Control flow action objects, such as \texttt{While} and \texttt{If}, contain sub-sequences of actions, resulting in a hierarchical data-structure similar to an abstract syntax tree.  This view of the compiler internals reveals that the metaphor of \emph{recording actions} is really an abstraction over the construction of program fragments.


\subsection{Backend Targets} \label{sec:design-backend}

\textbf{fault} supports a variety of open-source and commercial backend targets for running tests.  A target is responsible for consuming an action sequence, compiling it into a format compatible with the target runtime, and providing an API for invoking the runtime.  
Targets must also report the result of the test either by reading the exit code of running the process or processing the test output.

\subsubsection{Verilog Simulation Targets}
The \textbf{fault} compiler includes support for the open-source Verilog simulators \textbf{verilator}~\cite{snyder2004verilator} and \textbf{iverilog}~\cite{williams2006icarus}, plus three commercial simulators.  To compile \textbf{fault} programs to a \textbf{verilator} test bench, the backend lowers the action sequence into a \CC{} program that interacts with the software simulation object produced by the \textbf{verilator} compiler.  For \textbf{iverilog} and the commercial simulators, the backend lowers the action sequence into a SystemVerilog test bench that interacts with the test circuit through an \texttt{initial} block inside the top-level module.  One useful aspect of the SystemVerilog backend is its handling of variations in the feature support of target simulators.  For example, the commercial simulators use different commands for enabling waveform tracing and \textbf{iverilog} uses a non-standard API for interacting with files.  Constrained random inputs are generated using rejection or SMT~\cite{dutra2018smtsampler} sampling.




\subsubsection{CoSA}
The CoreIR Symbolic Analyzer (CoSA) is a solver-agnostic SMT-based hardware model checker~\cite{cosa}.  \textbf{fault}'s CoSA target relies on \textbf{magma}'s ability to compile Python circuit descriptions to CoreIR~\cite{coreir}, a hardware intermediate representation.  CoreIR's formal semantics are based on finite-state machines and the SMT theory of fixed-size bitvectors~\cite{BarFT-SMTLIB}. 
 \textbf{fault} action sequences are lowered into CoSA's custom explicit transition system format (ETS) and combined with the CoreIR representation of the circuit to produce a model.  CoSA allows the user to specify assumptions and properties, providing a straightforward lowering of \textbf{fault} assumptions and guarantees.

\subsubsection{SPICE}
In addition to being able to test designs with Verilog simulators, \textbf{fault} supports analog and mixed-signal simulators.  Compared to the traditional approach of maintaining separate implementations for digital and analog tests, this is a significantly easier way to write tests for mixed-signal circuits.
Basic actions such as \texttt{poke} and \texttt{expect} are supported in the SPICE simulation mode, but they are implemented quite differently than they are in Verilog-based tests.  Rather than emitting a sequential list of actions in an \texttt{initial} block, \textbf{fault} compiles \texttt{poke} actions into piecewise-linear (PWL) waveforms.  Other actions, such as \texttt{expect}, are implemented by post-processing the simulation data.

\subsubsection{Verilog-AMS}
For designs containing a mixture of SPICE and Verilog blocks, \textbf{fault} supports testing with a Verilog-AMS simulator.  This mode is more similar to running SystemVerilog-based tests than SPICE-based tests.
In particular, the test bench is implemented using a  top-level SystemVerilog module, meaning that a wide range of actions are supported including loops and conditionals.  This is a key benefit of using a Verilog-AMS simulator as opposed to a SPICE simulator.



\section{Evaluation}
To demonstrate \textbf{fault}'s capabilities, we evaluate the runtime performance of four different testing tasks from the domain of hardware verification.  Each task highlights the utility of \textbf{fault}'s portability by reusing the same source input across separate trials of different targets.  Due to licensing restrictions, we omit the name of the commercial simulators and replace them with a generic name.  The code to reproduce these experiments is available in the artifact.\footnote{\url{https://github.com/leonardt/fault\_artifact/blob/master/README.md}}  Each experiment involves at least one open-source simulator, but reproducing all the results requires access to commercial simulators.

\subsubsection{CGRA Processing Element Unit Tests}
To demonstrate the capability of \textbf{fault} as a tool for writing portable tests for digital verification, Figure~\ref{fig:lassen_unit_tests} reports the runtime performance of a subset of the \textbf{lassen} test suite.  \textbf{lassen}~\cite{lassen} is an open-source implementation of a CGRA processing element that contains a large suite of unit tests using \textbf{fault}.
Interestingly, we see comparable performance between \textbf{verilator} and \textbf{commercial simulator 1}, while \textbf{commercial simulator 2} is consistently $\sim$5x slower than the others.  One important property of the \textbf{lassen} test suite is that it generates a new test bench for each operation and input/output pair.  This stresses a simulator's ability to efficiently handle incremental changes, since each invocation involves a new top-level test bench file, but an unchanged design under test.

\begin{figure}
    \centering
    \setlength{\tabcolsep}{5pt}
\begin{tabular}{l c c c}
    Test                 & verilator & commercial sim 1  & commercial sim 2 \\
    \hline
    \verb|test_unsigned_binary| & 94.483  & 88.700 & 519.079 \\
    \verb|test_smult|           & 31.439  & 28.668 & 170.115 \\
    \verb|test_fp_binary_op|    & 104.117 & 91.878 & 571.759 \\
    \verb|test_stall|           & 10.424  & 9.629  & 56.458 \\
\end{tabular}
    \caption{Runtime (s) for unit tests of a CGRA processing element collected with a VM running on an Intel(R) Xeon(R) Silver 4214 CPU @ 2.20GHz with 256GB of RAM.}
    \label{fig:lassen_unit_tests}
\end{figure}

\subsubsection{SRAM Array}
To demonstrate the capability of \textbf{fault} as a tool for writing portable tests for analog and mixed-signal verification, we used \textbf{OpenRAM} to generate a 16x16 SRAM and then ran a randomized readback test of the design with SPICE, Verilog-AMS, and SystemVerilog simulators.
\textbf{OpenRAM}~\cite{7827670} is an open-source memory compiler that produces a SPICE netlist and Verilog model.

\begin{figure}
\centering
\subfloat[Runtime using a VM on an Intel(R) Xeon(R) CPU E5-2680 v4 @ 2.40GHz with 64GB of RAM.]{
\begin{tabular}[t]{l@{\hskip 0.3cm} l c}
Target & Simulator & Runtime (s) \\ [0.5ex] 
\hline
spice & ngspice & 117.660 \\ 
spice & comm sim 1 & 199.868 \\
spice & comm sim 2 & 98.043 \\
system-verilog & iverilog & 0.238 \\
system-verilog & comm sim 1 & 1.081 \\
system-verilog & comm sim 2 & 2.807 \\
verilog-ams & comm sim 1 & 228.405 \\
\end{tabular}
\label{fig:sram-runtime}
}
\hspace{0.75cm}
\subfloat[LoC for \textbf{fault} and language-specific implementations of the test.]{
\begin{tabular}[t]{l c}
\multicolumn{2}{c}{Lines of Code (LoC)} \\ [0.5ex] 
\hline
\textbf{fault} & 136 \\
\addlinespace[0.19cm]
\makecell[l]{Handwritten \\ SPICE} & 223 \\
\addlinespace[0.19cm]
\makecell[l]{Handwritten \\ SystemVerilog \\ and Verilog-AMS} & 189 \\
\end{tabular}
\label{fig:sram-loc}
}
\caption{Results for OpenRAM 16x16 SRAM randomized readback test.}
\end{figure}

The results shown in Figure~\ref{fig:sram-runtime} reveal two interesting trends.  First, as expected, SPICE simulations of the array were significantly slower than Verilog simulations (100-1000x).  Since \textbf{fault} allows the user to prototype tests with fast Verilog simulations, and then seamlessly switch to SPICE for signoff verification, our tool may reduce the latency in developing mixed-signal tests by orders of magnitude.
Second, even for simulations of the same type, there was significant variation in the runtime of different simulators.  SPICE simulation time varied by about 2x, while Verilog simulation time varied by about 10x.  One of the advantages of using \textbf{fault} is that it is easy to switch between simulators to find the one that works best for a particular scenario.

We also looked at the amount of human effort required to use \textbf{fault} to implement this test as compared to the traditional approach of writing separate testbenches for each simulation language.  Since ``human effort'' is subjective, we used lines of code as a rough metric, as measured from handwritten implementations of the same test in SystemVerilog, Verilog-AMS, and SPICE.  Figure~\ref{fig:sram-loc} shows the results of this experiment: the \textbf{fault}-based approach used 136 LoC as compared to 412 LoC for the traditional approach, a reduction of 3.02x.

\subsubsection{CGRA Integration Test Bench}
To observe how \textbf{fault} scales to more complex testing tasks, we report numbers for an integration test of the Stanford Garnet CGRA~\cite{garnet}.  This test generates an instance of the CGRA chip, runs a simulation that programs the chip for an image processing application, streams the input image data onto the chip, and streams the output image data to a file.  The output is compared to a reference software model.  Running the test took 232 minutes with the \textbf{verilator} target, 185 minutes with \textbf{commercial simulator 1}, and 221 minutes with \textbf{commercial simulator 2}.  Leveraging the portability of \textbf{fault}-based tests could save up to 47 minutes in testing time.  These results were collected using the same machine as the SRAM experiment (see Figure~\ref{fig:sram-runtime}).

\subsubsection{Unified Constrained Random and Formal}
To demonstrate the utility of the assume/guarantee interface as a unified abstraction for constrained random and formal verification,
we compared the runtime performance of using a constrained random target versus a formal model checker to verify the simple ALU property
shown in Section~\ref{sec:assume-guarantee}.  The first test evaluated the runtime performance of verifying correctness of the property on 100 constrained random inputs versus using a formal model checker.  The formal model checker provided a complete proof of correctness using interpolation-based model checking~\cite{interpolation} in 1.613~s, while constrained random verified 100 samples in 2.269~s (rejection sampling) and 2.799~s (SMT sampling).  The second test injected a bug into the ALU by swapping the opcodes for addition and subtraction. The model checker found a counterexample in 1.154~s with bounded model checking~\cite{bmc}, while constrained random failed in 2.947~s (rejection sampling) and 1.230~s (SMT sampling).  In both cases the model checker was at least as fast as the constrained random equivalent while providing better coverage in the case of no bug.  These results were collected using a MacBook Pro (13-in 2017, 4 Thunderbolt, macOS 10.15.2), with a 3.5 GHz Dual-Core Intel i7 CPU, and 16 GB RAM.




\section{Related Work}

Prior work has leveraged using a generic API to Verilog simulators to build portability into testing infrastructures.
The \textbf{ChiselTest} library~\cite{chisel-testers2} and \textbf{cocotb}~\cite{cocotb} provide this capability for Scala and Python respectively.
Using a generic API offers many of the same advantages with regards to test portability, simplicity, and automation, but the lack of
multi-stage execution limits the application to more diverse backend targets such as SPICE simulations and formal model checkers.
However, because these libraries interact with the simulator directly, they do allow user code to immediately respond to the
simulator state, enabling interactive debugging through the host language.  \textbf{cocotb} also presents a coroutine abstraction that
naturally models the concurrency found in hardware simulation.  Future work could investigate using cocotb as a runtime target 
for \textbf{fault}'s frontend, enabling a similar concurrent, interactive style of testing.  Another interesting avenue of work would be 
to extend \textbf{fault}'s backend targets to support lowering \textbf{cocotb}'s coroutine abstraction.


\section{Conclusion}
The ethos of \textbf{fault} is to enable the construction of flexible, portable test components that are simple to integrate and scale for testing complex applications.  The ability to metaprogram test components is essential for enabling verification teams to match the productivity of design teams using generators.  
\textbf{fault}'s portability enables teams to easily transition to different tools for different use cases, and enables the proliferation of reusable verification libraries that are applicable in a diverse set of tooling environments.  

While \textbf{fault} has already demonstrated utility to design teams in academia and industry, there remains a bright future filled with opportunity to improve the system.  Extending the assume/guarantee interface to support temporal properties/constraints and leverage compositional reasoning~\cite{DBLP:conf/lics/ClarkeLM89} is essential for scaling the approach to more complex systems.  Adding concurrent programming abstractions such as coroutines are essential for capturing the common patterns used in the testing of parallel hardware.  Using a deep embedding architecture could significantly improve the performance of generating \textbf{fault} test benches.


\subsubsection*{Funding} The authors would like to thank the DARPA DSSoC (FA8650-18-2-7861) and POSH (FA8650-18-2-7854) programs, the Stanford AHA and SystemX affiliates, Intel’s Agile ISTC, the Hertz Foundation Fellowship, and the Stanford Graduate Fellowship for supporting this work.

\bibliographystyle{splncs04}
\bibliography{main}
\end{document}